\documentstyle[11pt,aaspp4,psfig,flushrt]{article}
\begin{document}

\def\sax{SAX\,J1808.4--3658}

\lefthead{Psaltis \& Chakrabarty}
\righthead{Disk-Magnetosphere Interaction in \sax}

\title{The Disk-Magnetosphere Interaction in the
Accretion-Powered Millisecond Pulsar \sax}

\author{Dimitrios Psaltis}
\affil{\footnotesize Harvard-Smithsonian Center for
Astrophysics, 60 Garden St., Cambridge, MA 02138; 
dpsaltis@cfa.harvard.edu}

\and

\author{Deepto Chakrabarty}
\affil{\footnotesize Department of Physics and Center for Space
Research, Massachusetts Institute of
Technology,\\ Cambridge, MA 02139; deepto@space.mit.edu}

\vspace*{0.3cm}

\begin{center}
To appear in {\em The Astrophysical Journal\/.}
\end{center}

\begin{abstract}
The recent discovery of the first known accretion-powered millisecond
pulsar with the {\em Rossi X-Ray Timing Explorer} provides the first direct
probe of the interaction of an accretion disk with the magnetic field of a
weakly magnetic ($B\lesssim 10^{10}$~G) neutron star. We demonstrate that
the presence of coherent pulsations from a weakly magnetic neutron star
over a wide range of accretion rates places strong constraints on models
of the disk-magnetosphere interaction. We argue that the simple $\dot
M^{3/7}$ scaling law for the Keplerian frequency at the magnetic
interaction radius, widely used to model disk accretion onto magnetic
stars, is not consistent with observations of \sax\ for most proposed
equations of state for stable neutron stars.  We show that the usually
neglected effects of multipole magnetic moments, radiation drag forces,
and general relativity must be considered when modeling such weakly
magnetic systems.  

Using only very general assumptions, we obtain a robust
estimate of $\mu\simeq (1-10)\times 10^{26}$~G~cm$^3$ for the dipole
magnetic moment of \sax, implying a surface dipole field of
$\sim10^8-10^9$~G at the stellar equator. We therefore infer that after
the end of its accretion phase, this source will become a normal
millisecond radio pulsar.  Finally, we compare the physical properties of 
this pulsar to those of the non-pulsing, weakly magnetic neutron stars in 
low-mass X-ray binaries and argue that the absence of coherent pulsations 
from the latter does not necessarily imply that these neutron stars 
have significantly different magnetic field strengths from \sax.
\end{abstract}

\keywords{accretion, accretion disks --- pulsars: individual: \sax\
--- stars: neutron --- X-rays: stars}

\newpage

\section{INTRODUCTION}

The basic framework for understanding accretion-powered pulsars
emerged soon after their discovery (Giacconi et al.\markcite{Getal71}
1971). These systems are rotating neutron stars accreting matter from
a binary companion, with magnetic fields strong enough to disrupt the
accretion flow above the stellar surface (Pringle \&
Rees\markcite{PR72} 1972; Davidson \& Ostriker\markcite{DO73} 1973;
Lamb, Pethick, \& Pines\markcite{LPP} 1973). When threaded by the
stellar magnetic field, the accreting gas is brought into corotation
with the star and is channeled along field lines to the polar caps,
releasing its potential and kinetic energy mostly in X-rays. The
rotation of these hot spots through our line of sight produces X-ray
pulses at the spin frequency of the neutron star.  Most of the
$\approx 50$ accretion-powered pulsars have spin frequencies and
inferred dipole magnetic field strengths in the range $\sim$$1-10^3$~s
and $\sim$$10^{11}-10^{13}$~G, respectively (White, Nagase, \&
Parmar\markcite{WNP95} 1995).

In strongly magnetic ($B\gtrsim 10^{11}$~G) accreting neutron stars,
models for the disk-magnetosphere interaction based on this framework (see
Ghosh \& Lamb\markcite{GL91} 1991 for a review) are generally consistent
with the accretion torque behavior of the Be/X-ray pulsar transients, as
observed by the {\em Compton}/BATSE all-sky monitor (see Bildsten et
al.\markcite{Betal97} 1997 and references therein).  The Be/X-ray pulsar
transients allow for a direct test of the predicted scaling of torque
with accretion rate, since they span a wide range of accretion rates
during their outbursts. However, the same models can only account for the
bimodal torque reversals observed in several persistent accreting pulsars
if additional assumptions are introduced (Chakrabarty et
al.\markcite{Cetal97a}\markcite{Cetal97b} 1997a, 1997b; Bildsten et al.\
1997).  Examples of suitable assumptions are a bimodal distribution of the
mass transfer rate onto the pulsar, or a bimodal dependence on accretion
rate of the orientation of the disk (Nelson et al.\markcite{Netal97} 1997;
van Kerkwijk et al.\markcite{vKetal98} 1998), of the orbital angular
velocity of the accreting gas (Yi \& Wheeler\markcite{YW98} 1998), or even
of the strength and orientation of any magnetic field produced in the disk
(Torkelsson\markcite{T98} 1998). 

In weakly-magnetic ($B\lesssim 10^{10}$~G) accreting neutron stars,
models of the disk-magnetosphere interaction have only been tested
indirectly so far.  Most neutron stars in low-mass X-ray binaries (LMXBs)
show no periodic oscillations in their persistent emission. They have
dipole magnetic fields $\lesssim 10^{10}$~G, as inferred from their
bursting (Lewin, van Paradijs, \& Taam\markcite{LPT95} 1995) and rapid
variability behavior (see van der Klis\markcite{vdK98} 1998 for a review)
as well as from their X-ray spectra (Psaltis \& Lamb\markcite{PL98}
1998).  Their power-density spectra show various types of quasi-periodic
oscillations (QPOs), and in particular the most luminous of the
neutron-star LMXBs show $\sim 15-60$~Hz QPOs. These are called
horizontal-branch oscillations (HBOs) and have frequencies that increase
with mass accretion rate (van der Klis et al.\markcite{vdk1985} 1985; van
der Klis\markcite{vdK89} 1989). They have been interpreted as occurring at
the beat frequency between the Keplerian frequency at the magnetic
interaction radius (where the stellar magnetic field strongly interacts
with the accretion disk) and the neutron star spin frequency (Alpar \&
Shaham\markcite{AS85} 1985; Lamb et al.\markcite{Letal85} 1985). Models
of the inner accretion disk and the disk-magnetosphere interaction can
account for the observed scaling of HBO frequency with accretion rate if
the neutron stars in these systems are near their magnetic spin
equilibrium and if the inner accretion disks are radiation-pressure
dominated (Psaltis et al.\markcite{Petal99} 1999b). 

The discovery of $\sim 200-1200$~Hz QPOs (often occurring in pairs;
hereafter kHz QPOs; see van der Klis 1998 and references therein) in the
X-ray flux of many neutron-star LMXBs and the identification of the
higher-frequency QPO with a Keplerian orbital frequency in the accretion
disk (van der Klis et al.\markcite{vdk96} 1996; Strohmayer et
al.\markcite{Setal96} 1996; Miller, Lamb, \& Psaltis\markcite{MLP98}
1998) has introduced additional complications in studying the
disk-magnetosphere interaction using HBO observations (see also Psaltis,
Belloni, \& van der Klis\markcite{pbk99} 1999a). The magnetic interaction
radius inferred from the HBO frequencies is larger than the disk radius
responsible for the higher-frequency kHz QPO. Therefore, if the
higher-frequency kHz QPO is a Keplerian orbital frequency in the disk,
then the magnetospheric beat-frequency model for the HBO can be valid
only if a non-negligible fraction of the disk plasma is not threaded by
the stellar magnetic field but remains in the disk plane inside the
interaction radius (see Miller et al.\ [1998] and Psaltis et al.\ [1999b]
for a discussion).  This requirement has not yet been addressed in any
theoretical model for the disk-magnetosphere interaction and therefore 
cannot be tested directly. 

The recent discovery of the first weakly-magnetic accretion-powered
pulsar with the {\em Rossi X-ray Timing Explorer (RXTE)\/} allows the
first direct test of disk-magnetosphere interaction models in the
weak-field limit.  \sax\ is a transient X-ray source that shows type~I
X-ray bursts (in't Zand et al.\markcite{Zetal98} 1998) and coherent
401~Hz X-ray pulsations (Wijnands \& van der Klis\markcite{WK98a} 1998a),
and is a member of a low-mass binary in a 2-hour orbit (Chakrabarty \&
Morgan\markcite{CM98} 1998).  Its distance is $\approx$4~kpc, as inferred
from flat-topped type~I X-ray bursts that are thought to be Eddington
limited (in 't Zand et al.\ 1998), and its luminosity varied by
$\gtrsim$2 orders of magnitude during the 1998 April/May outburst (Cui,
Morgan, \& Titarchuk\markcite{CMT98} 1998). 

In this paper we study the disk-magnetosphere interaction in \sax. In
\S2, we compare the predictions of theoretical models for the inner
accretion disk and the disk-magnetosphere interaction with observations
of \sax\ and, in \S3, we infer the magnetic field strength of the pulsar.
In \S4, we discuss our results and their implications for
disk-magnetosphere interaction models and compare \sax\ to the
millisecond radio pulsars and the non-pulsing neutron stars in LMXBs. 

\section{DISK-MAGNETOSPHERE INTERACTION}

\subsection{Assumptions and Formalism}

Throughout this paper we assume that most of the accreting gas around
\sax\ is confined in a geometrically thin accretion disk before
interacting with the pulsar magnetic field.  This assumption is supported
by the transient nature of the source, which can be understood in terms of
dwarf-nova--like accretion disk instabilities (see Chakrabarty \& Morgan
1998).  We also neglect the effect of wind mass loss from the inner
accretion disk and of radiation drag forces, as well as all general
relativistic effects. 

The radius $r_0$ at which magnetic stresses remove the angular momentum of
the disk flow and disrupt it can be estimated by balancing
the magnetic and material stresses, (Ghosh \&
Lamb\markcite{GL78}\markcite{GL79a} 1978, 1979a)
 \begin{equation}
 \frac{B_{\rm p} B_{\rm \phi}}{4\pi} 4\pi r_0^2 \Delta 
 r_0 = \dot{M} \Omega r_0^2\;,
 \label{bal}
 \end{equation}
 where $B_{\rm p}$ and $B_{\rm \phi}$ are the poloidal and toroidal
components of the magnetic field, $\Delta r_0$ is the radial width of the
interaction region, $\dot{M}$ is the mass transfer rate through the inner
disk, and $\Omega(r)$ is the angular velocity of the gas at radius $r$. 
Assuming that the poloidal magnetic field is dipolar with magnetic moment
$\mu$ and that the accretion flow is Keplerian, we obtain
 \begin{equation}
   r_0= \gamma_{\rm B}^{2/7}
 \left(\frac{\mu^4}{GM\dot{M}^2}\right)^{1/7}\;,
 \label{gen}
 \end{equation}
 where $\gamma_{\rm B}\equiv (B_\phi/B_p)(\Delta r_0/r_0)$, $M$ is the
mass of the neutron star, and $G$ is the gravitational constant. 

The boundary layer parameter $\gamma_{\rm B}$ in equation~(\ref{gen}) depends
in general on all the other physical quantities. Hence, different models for
the inner accretion disk's environment and the disk-magnetosphere interaction
generally predict different coefficients and scalings (Ghosh \& Lamb 1992;
see also Ghosh \& Lamb 1978, 1979a; Wang\markcite{W96} 1996;
Ghosh\markcite{G96} 1996).  However, if the disk-magnetosphere interaction
takes place in a region of the accretion disk where all physical quantities
have a power-law dependence on radius (as is the case for most accretion disk
models away from the stellar surface) and equation~(\ref{bal}) describes
angular momentum balance in the interaction region, then the Keplerian
orbital frequency at $r_0$ can be written as
 \begin{equation} 
  \nu_{\rm 0} \simeq \nu_{\rm K,0}
  \left(\frac{M}{M_\odot}\right)^{\gamma}
  \left(\frac{\mu}{10^{27}~\mbox{G\,cm}^{3}}\right)^{\beta}
  \left(\frac{\dot{M}}{\dot{M}_{\rm E}}\right)^{\alpha}\;, 
 \label{mag}
 \end{equation}
 where $M_\odot$ is the solar mass; $\dot{M}_{\rm E}=2\times
10^{-8}\,M_\odot$~yr$^{-1}$ is the Eddington critical accretion rate (at which
the outward radiation forces in a spherically symmetric hydrogen flow balance
gravity); and $\alpha$, $\beta$, $\gamma$, and $\nu_{\rm K,0}$ are
parameters which depend upon the nature of the inner accretion disk. 
For our study, we used the parameters from four different inner disk
models (see Table~1).  We chose these models both because they span a
wide range of physical conditions in the inner disk (one- and
two-temperature plasmas, and gas and radiation-pressure--dominated
flows) and a range of radiation processes, and because detailed
calculations of the interaction with a stellar magnetic field have
been performed for these models (see Table~1 and Ghosh \& Lamb 1992
for a discussion of the physical processes involved in these
models). 

\subsection{Limits from the Detection of Coherent Pulsations}

We can use the presence of coherent X-ray pulsations in \sax\ over a
wide range of mass accretion rate to test scaling (\ref{mag}) in the
weak magnetic field regime.  For a rotating neutron star to appear as
an accretion-powered pulsar, the stellar magnetic field must be strong
enough to disrupt the Keplerian disk flow above the stellar
surface\footnote{Note that this is a necessary but not a sufficient
condition: the X-ray brightness at infinity may not be modulated with
any detectable amplitude if, e.g., the spin of the neutron star is
perfectly aligned to its magnetic moment, or the neutron star is
surrounded by a scattering medium that attenuates the pulsations
(Brainerd \& Lamb 1986; Kylafis \& Phinney 1989).}.  Note that the
orbital frequency $\nu_0$ at the interaction radius $r_0$ increases with 
mass accretion rate ($\alpha>0$).  Therefore, the stellar magnetic field
must be strong enough to disrupt the disk flow at radii
larger than the neutron star radius $R$ at the {\em maximum\/} mass
accretion rate $\dot{M}_{\rm max}$ for which coherent pulsations were
detected. This leads to a lower limit on the magnetic dipole moment,
 \begin{equation}
 \mu \ge 10^{27}
 \left(\frac{4\pi^2\nu_{\rm K,0}^2 R^3}
   {GM_\odot}\right)^{-1/2\beta}
 \left(\frac{M}{M_\odot}\right)^{(1-2\gamma)/2\beta}
 \left(\frac{\dot{M}_{\rm max}}{\dot{M}_{\rm E}}\right)
   ^{-\alpha/\beta}~\mbox{G~cm}^3\;.
 \label{lower}
 \end{equation}
 At the same time, the stellar magnetic field must be weak enough that
accretion is not centrifugally inhibited at the {\em minimum\/} mass
accretion rate $\dot{M}_{\rm min}$ for which coherent pulsations were
detected. Therefore, the orbital frequency $\nu_0$ at this accretion
rate must be larger than the spin-frequency of the neutron star
$\nu_{\rm s}$. This leads to an upper limit on the magnetic dipole
moment,
 \begin{equation}
 \mu \le 10^{27}
 \left(\frac{\nu_{\rm s}}{\nu_{\rm K,0}}\right)^{1/\beta}
 \left(\frac{M}{M_\odot}\right)^{-\gamma/\beta}
 \left(\frac{\dot{M}_{\rm min}}{\dot{M}_{\rm E}}\right)
    ^{-\alpha/\beta}~\mbox{G~cm}^3\;.
 \label{upper}
 \end{equation}

Of course, X-ray brightness modulations at the pulsar spin frequency may be
possible even if one of the above two requirements is not met. For
example, azimuthal variations in the efficiency of processes dependent on
magnetic field strength (such as cyclotron emission, absorption, and
resonant scattering) may be strong enough to produce X-ray pulsations,
even if the magnetic field of the neutron star is dynamically unimportant
and the accretion disk extends down to the stellar surface.  However, if
at any point during the outburst either the disk had reached the stellar
surface or accretion had been centrifugally inhibited, then either the
X-ray spectrum or the pulse amplitude should have changed abruptly. 
Instead, the X-ray spectrum of \sax\ was found to be remarkably stable
(Gilfanov et al.\markcite{Getal98} 1998; Heindl \& Smith\markcite{HS98}
1998), and the pulse amplitude to increase only slightly from 4\% to
7\% as the inferred luminosity declined by more than two orders of
magnitude during the outburst (Cui et al.\ 1998). 
 
\begin{figure}[h]
\begin{minipage}[b]{10.5cm}
\centerline{
\psfig{file=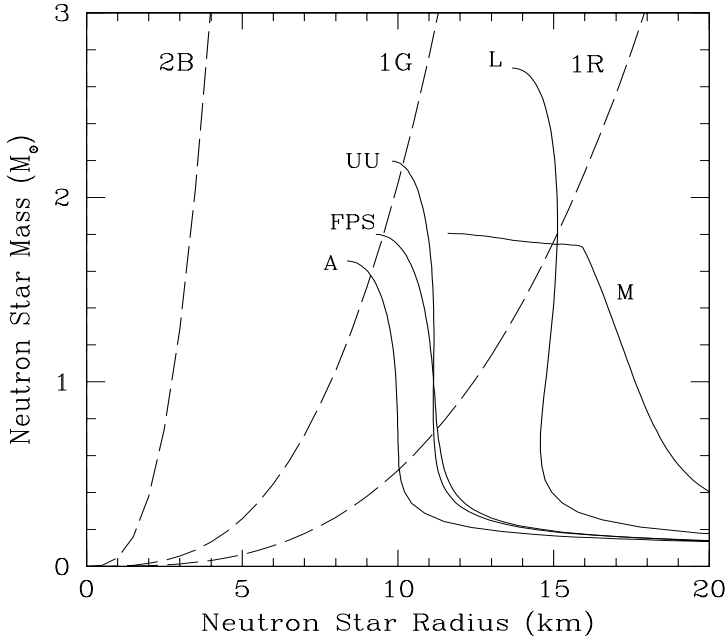,angle=0,height=8.8truecm,width=10truecm}}
\end{minipage}
\begin{minipage}[b]{5.0cm}
\figcaption[]
 {\footnotesize Constraints on the neutron star mass and radius in \sax.
The solid lines represent mass-radius relations of stable neutron stars
for various proposed equations of state (Cook et al. 1994).  The dashed
lines represent the limits imposed by various models for the inner
accretion disk (Table~1).  The allowed region for each inner disk model is
to the left of the corresponding dashed curve.  The limiting curve for the
2S model essentially lies on the $R=0$ axis on this scale, and is not
plotted. }
 \vspace*{1.3cm}
 \end{minipage}
\end{figure}

Combining constraints~(\ref{lower}) and (\ref{upper}) we obtain
 \begin{equation}
 \frac{M}{M_\odot} > 0.047
 \left(\frac{\nu_{\rm s}}{401~\mbox{Hz}}\right)^2
 \left(\frac{\dot{M}_{\rm max}}{\dot{M}_{\rm min}}
    \right)^{2\alpha} 
 \left(\frac{R}{10~\mbox{km}}\right)^3\;.
 \label{MRlim}
 \end{equation}
 Given the observed spin frequency and range of accretion rates at
which coherent pulsations are detected, inequality~(\ref{MRlim})
defines a limiting curve in the $M$-$R$ parameter space for neutron
stars. The scaling of this curve is cubic in $R$, independent of the
details of the disk-magnetosphere interaction model, with a
coefficient that depends on a single model parameter\footnote{While
 this work was in progress, we learned of a paper by Burderi \&
 King\markcite{BK98} (1998) in which a similar argument is 
 used to constrain $R$ by assuming a particular disk-magnetosphere
 interaction model (essentially the 1G model) and hence a specific
 value for $\alpha$.  However, the applicability of the 1G model to
 \sax\ is dubious, as we demonstrate here.  In contrast, we use
 inequality~(\ref{MRlim}) to constrain disk-magnetosphere interaction
 models and hence $\alpha$.}.  As  
mentioned above, the X-ray spectrum of \sax\ was found to be
remarkably stable during the outburst, so we can assume here that the
ratio $\dot{M}_{\rm max}/\dot{M}_{\rm min}$ scales as the ratio of
the corresponding countrates in a given X-ray bandpass, independent of
(unknown) bolometric corrections.  For the 1998 April/May outburst of
\sax, this ratio was $\approx 130$ (Cui et al.\ 1998).

Figure~1 compares the $M$-$R$ relations for several proposed equations of
state for stable neutron stars (see Cook, Shapiro, \&
Teukolsky\markcite{CST94} 1994) to the limit~(\ref{MRlim}) for the four
previously mentioned inner disk models applied to \sax.  Only the limiting
curve for the 1R model, for which $\alpha\simeq 0.2$, is consistent with
many proposed equations of state. In a thin accretion disk, radiation
pressure dominates over gas pressure at radii (see Treves,
Maraschi, \& Abramowicz 1988)
 \begin{equation}
 \frac{r}{R}\ll 0.5 \alpha^{0.1}
 \left(\frac{\eta}{500}\right)
 \left(\frac{M}{M_\odot}\right)^{0.42}
 \left(\frac{R}{10~\mbox{km}}\right)^{-0.24}
 \left(\frac{\dot{M}}{10^{-4}~\dot{M}_{\rm E}}\right)^{0.76}\;,
 \end{equation}
where $\eta$ is a parameter in the range $10^2$--$10^3$ and
$\alpha\lesssim 1$ is the Shakura-Sunyaev viscosity parameter. During
the decline phase of the outburst, the mass accretion rate was so
small that $r/R\ll1$ and the inner accretion disk in \sax\ was
certainly not radiation-pressure dominated.  Whether or not the disk was
radiation-pressure dominated at the peak of the outburst is not clear
given the uncertainty in the model parameter $\eta$. However, it is
very unlikely that there was a transition between a radiation-pressure
and a gas-pressure--dominated inner accretion disk during the
outburst, given the stability of the X-ray spectrum of the source and
of the pulsed fraction of the emission (see discussion below).
Radiation-pressure--dominated models for the inner accretion disk,
like the one shown in Figure~1, are therefore not applicable to \sax\
since its inferred mass accretion rate is significantly sub-Eddington
(in't Zand et al.\ 1998). 

   From the three gas-pressure dominated models, the two-temperature ones
(2B and 2S) are inconsistent with any of the proposed equations of
state and are thus ruled out for \sax. The one-temperature
gas-pressure dominated model (1G) is {\em barely\,} consistent with a
very restricted range of $M$ and $R$ allowed by only a few equations
of state.  However, the maximum and minimum accretion rates at which
{\em RXTE\/} detected coherent pulsations from \sax\ were determined
by the peak luminosity of the outburst and the instrumental
sensitivity, respectively, which are obviously unrelated to the
limiting equalities in (\ref{lower}) and (\ref{upper}).  This makes it
quite unlikely that the actual mass and radius of \sax\ lies very
close to the limiting curve 1G in Figure~1.  Therefore, either a
remarkable coincidence has occurred or (more probably) the 1G model is
also ruled out for \sax.  This is particularly interesting because the
dependence of $\nu_0$ on $\dot M$ predicted by the 1G model is very
similar to the simple $\nu_0\propto \dot{M}^{3/7}$ scaling law derived
from equation~(\ref{gen}) for constant $\gamma_{\rm B}$.  This simple
scaling law is widely used to describe the disk-magnetosphere
interaction in accreting magnetic stars (and is in fact generally
considered the standard model).  Figure~1 implies that the scaling of
$\nu_0$ with $\dot M$ for \sax\ should be weaker than predicted by
this simple scaling law.  In particular, we find that $\alpha$ must be
$< 0.4$, and hence $\gamma_{\rm B}$ must be a function of $\dot M$.

\subsection{Predicted Limits on the Accretion Torque in \sax}

An independent test of scaling (\ref{mag}) in the weak magnetic field
regime is possible if the spin frequency derivative $\dot{\nu}_{\rm s}$ in 
\sax\ due to accretion torques can be measured.  For a Keplerian disk flow
terminated at a radius $r_0$, the accretion torque on the neutron star
is given by
 \begin{equation}
 2\pi I\dot{\nu}_{\rm s} =
 \eta \dot{M}\left(GMr_0\right)^{1/2}\;,
 \label{torque}
 \end{equation}
 where $I$ is the neutron star moment of inertia and $\eta$ is a
dimensionless quantity in which all the physics of the
disk-magnetosphere interaction is parametrized (Ghosh \& Lamb
1979b). For a prograde accretion disk, $\eta$ is positive and a strong
function of both the magnetic field strength and the mass accretion
rate (Ghosh \& Lamb 1979b). For a retrograde accretion disk $\eta$ is
negative and has an absolute value of order unity (Daumerie 1996).

Applying constraints~(\ref{lower}) and (\ref{upper}) to
equation~(\ref{torque}), we obtain an upper and a lower limit on the spin
frequency derivative for \sax\ predicted by accretion torque theory,
 \begin{eqnarray}
 \dot{\nu}_{\rm s} &\le& 1.8\times 10^{-13}\eta
 \left(\frac{I/10^{45}~\mbox{g\,cm}^2}
   {M/M_\odot}\right)^{-1}
 \left(\frac{\nu_{\rm s}}{401~\mbox{Hz}}\right)^{-1/3}
 \nonumber\\
 & & \qquad\qquad
 \left(\frac{R_{\rm s}}{10~\mbox{km}}\right)
 \left(\frac{M}{M_\odot}\right)^{-1/3}
 \left(\frac{\dot{M}_{\rm max}}{0.06~\dot{M}_{\rm E}}
    \right)
 \left[\frac{(\dot{M}_{\rm min}/\dot{M}_{\rm max})^{\alpha/3}}
    {130^{1/7}}\right]
 \label{upperdot}
  \end{eqnarray}
 and
  \begin{equation}
 \dot{\nu}_{\rm s} \ge 1.0\times 10^{-13}\eta
 \left(\frac{I/10^{45}~\mbox{g\,cm}^2}
   {M/M_\odot}\right)^{-1}
 \left(\frac{R_{\rm s}}{10~\mbox{km}}\right)^{3/2}
 \left(\frac{M}{M_\odot}\right)^{-1/2}
 \left(\frac{\dot{M}_{\rm max}}{0.06~\dot{M}_{\rm E}}
    \right)\;.
 \label{lowerdot}
  \end{equation}
 The maximum observed 2--30~keV flux of \sax\ at the peak of the outburst
was $\simeq 3\times 10^9$~erg~cm$^{-2}$~s$^{-1}$ (Cui et al.\ 1998),
which corresponds to a 2--30~keV luminosity of $\simeq 0.03(\Omega_{\rm
e}/4\pi) L_{\rm E}$, where $\Omega_{\rm e}$ is the solid angle of
emission and $L_{\rm E}=2\times 10^{38}$~erg~s$^{-1}$ is the Eddington
critical luminosity that corresponds to $\dot{M}_{\rm E}$. However, the
photon spectrum of \sax\ over most of the {\em RXTE\/} 2--100~keV
bandpass is well described by a power-law of the form $dN/dE\propto
E^{-2}$, where $E$ is the photon energy.  The energy spectrum is thus
very flat, $\nu L_\nu\propto E^0$. As a result, the bolometric luminosity of
the source must be larger than estimated above and is probably about
twice the 2--30~keV luminosity, given the expected upper- and
lower-energy cut-offs of the spectrum (see also Heindl \& Smith 1998 for
spectral fits over the entire {\em RXTE\/} bandpass that show this
effect).  We therefore adopt $\dot{M}_{\rm max}=0.06\dot{M}_{\rm E}$. 

Given a measurement of the spin frequency derivative in \sax, we could
use relations (\ref{upperdot}) and (\ref{lowerdot}) to place
additional constraints on $\alpha$ for inner disk models or $M$ and
$R$ for neutron stars, similar to the discussion above for the
presence of coherent pulsations (Figure~1).  However, variations of
the spin frequency were not detected during the peak of the 1998
April/May outburst of \sax, leading to an upper limit on the spin
frequency derivative of $\vert \dot{\nu}_{\rm s}\vert < 7\times
10^{-13}$~Hz~s$^{-1}$ (Chakrabarty \& Morgan 1998).  This measured
upper limit is consistent with our predicted bounds on the spin
frequency derivative~(\ref{upperdot}) and (\ref{lowerdot}) if
$\eta\lesssim 7$. 

\section{THE MAGNETIC FIELD STRENGTH OF \sax}

In their discovery paper for \sax, Wijnands \& van der Klis (1998a)
applied ``standard'' magnetospheric disk accretion theory (essentially
the 1G model discussed in \S2) and used the absence of centrifugal
inhibition of accretion during the decline of the 1998 outburst to
infer that $B\lesssim (2-6)\times 10^8$~G.  As the outburst declined,
pulsations continued to be detected, leading to revised upper limits
on the magnetic field strength [$B\lesssim (0.4-1.3)\times 10^8$~G,
Cui et al. 1998; $B\lesssim$~few\ $\times 10^7$~G, Gilfanov et
al. 1998].  However, as we showed in \S2.2, simple scaling arguments
or models of the form~(\ref{mag}) for the asymptotic regions of
gas-pressure--dominated accretion disks cannot easily account for the
coherent pulsations observed from \sax\ throughout its 1998
outburst.  Therefore, the previous estimates of the magnetic field
strength of \sax\, which are based on similar scalings, are not valid.
A more careful calculation is necessary to infer the magnetic field
strength in this system.  Because no spin frequency derivative has
been detected yet from this source (Chakrabarty \& Morgan 1998), 
we cannot uniquely determine the stellar magnetic field strength, but
can only constrain it.

We begin with expression~(\ref{gen}) for the interaction radius, which
is derived directly from the angular momentum equation~(\ref{bal}),
and impose the same constraints that led to equations~(\ref{lower})
and (\ref{upper}). We set the maximum and minimum mass accretion rates
at which coherent pulsations were detected to $\dot{M}_{\rm max}\ge
0.06\dot{M}_{\rm E}$ and $\dot{M}_{\rm min}\le \dot{M}_{\rm max}/130$
as inferred from observations (see Cui et al.\ 1998 and \S2.3). We
assume a neutron star radius in the 10--15~km range, as predicted by
current equations of state (see Fig.\,1).  We also assume a neutron
star mass in the 1.4--2.3~$M_\odot$ range, consistent with the
inferred masses of recycled millisecond pulsars (Thorsett \&
Chakrabarty\markcite{TC99} 1999), the upper limits on neutron star
masses in LMXBs inferred from kHz QPO observations (Miller et al.\
1998; see also Zhang, Strohmayer, \& Swank\markcite{ZSS97} 1997), and
the possible neutron star mass measurement in 4U~1820$-$30 (Zhang et
al.\markcite{Zetal98a} 1998a).  

The limits on the magnetic field strength of \sax\ also depend upon
the allowed range for the parameter $\gamma_{\rm B}$, which in turn
depends upon the fractional width $\Delta r_0/r_0$ of the boundary
layer and the magnetic pitch $B_\phi/B_p$ within this layer (see Ghosh
\& Lamb 1991 for a detailed discussion).  The fractional width should
be significantly smaller than unity in order for the boundary layer
approximation~(\ref{gen}) to be valid. Assuming a dipolar poloidal
field and neglecting mass loss from the disk as well as toroidal
screening currents in both the disk and the magnetosphere, one finds
$\Delta r_0/r_0\simeq 0.3$ (Ghosh \& Lamb 1991; Daumerie
1996). However, relaxing any of these assumptions leads to a
significantly narrower boundary layer.  We therefore conservatively
adopt $0.01\lesssim \Delta r/r_0 < 1$.

The toroidal component $B_\phi$ of the magnetic field is produced by
the differential rotation of gas in the accretion disk with respect to
the stellar spin.  The magnetic energy stored in the magnetosphere
increases with the twisting of the field lines (see Zylstra 1988 and
references therein). There is an upper bound on this energy (and hence
on the twisting of the field lines and the magnetic pitch) for a
semi-infinite space in which the magnetic field strength goes to zero
at infinity (Aly\markcite{1984}\markcite{1991} 1984, 1991; Zylstra
1988).  It has been argued that, above this upper bound, the only
existing configurations are those with field lines that close at
infinity (i.e., open field lines; Aly\markcite{1984}\markcite{1991}
1984, 1991).  Simple estimates of the maximum twisting of the field
lines (see Ghosh \& Lamb 1991 and reference therein) as well as
detailed numerical calculations of the magnetospheric structure
(Zylstra 1988) lead to maximum values of $B_\phi/B_p \sim 1$.
Combining this with our constraints on $\Delta r_0/r_0$, we therefore
adopt $0.01\lesssim \gamma_{\rm B}(\dot M) \lesssim 1$.

Applying all these constraints to \sax, we obtain
 \begin{equation}
 \mu \gtrsim 0.3\times 10^{26} 
 \left[\gamma_{\rm B}(\dot{M}_{\rm max})\right]^{-1/2}
 \left(\frac{M}{1.4 M_\odot}\right)^{1/4}
 \left(\frac{R_{\rm s}}{10~\mbox{km}}\right)^{9/4}
 \left(\frac{\dot{M}_{\rm max}}{0.06 \dot{M}_{\rm E}}
    \right)^{1/2}~\mbox{G~cm}^{3}
 \end{equation}
and
 \begin{eqnarray}
 \mu & \lesssim & 10\times 10^{26} 
 \left[\frac{\gamma_{\rm B}(\dot{M}_{\rm min})}{0.01}\right]^{-1/2}
 \left(\frac{M}{2.3 M_\odot}\right)^{5/6}
 \left(\frac{R_{\rm s}}{15~\mbox{km}}\right)^{1/2}\nonumber\\
 & & \qquad\qquad\qquad
 \left(\frac{\nu_{\rm s}}{401~\mbox{Hz}}\right)^{-7/6}
 \left(\frac{\dot{M}_{\rm min}}{4.6\times 10^{-4} 
    \dot{M}_{\rm E}}\right)^{1/2}~\mbox{G~cm}^{3}\;,
 \label{muup}
 \end{eqnarray}
 which correspond to a dipolar magnetic field of a few times $10^8$~G at
the stellar pole.  Figure~2 shows the limits on the magnetic dipole moment
as a function of the (unknown) parameter $\gamma_{\rm B}(\dot M)$.  

\begin{figure}[t]
\begin{minipage}[b]{10.9cm}
\centerline{
\psfig{file=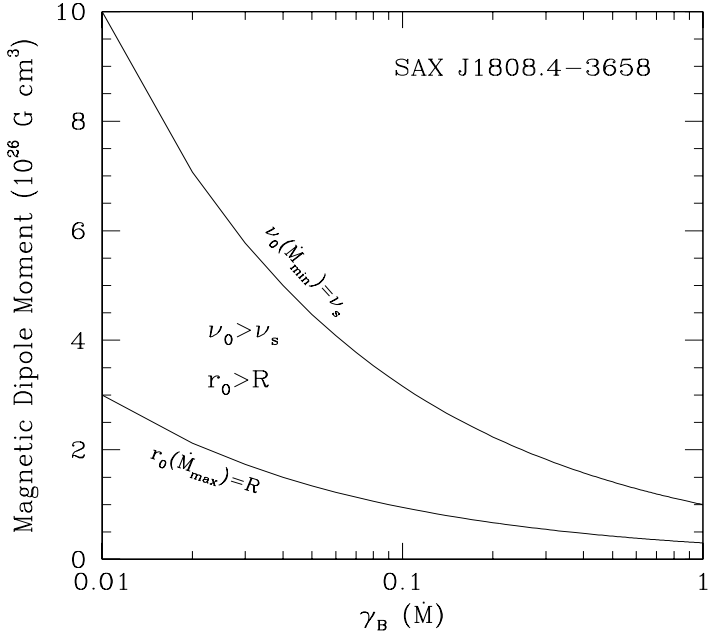,angle=0,height=8.8truecm,width=10truecm}}
\end{minipage}
\begin{minipage}[b]{5.0cm}
\figcaption[]  
 {\footnotesize Constraints on the magnetic dipole moment $\mu$ of \sax\
as a function of boundary layer parameter $\gamma_{\rm B}(\dot M)$.  The
upper limit on $\mu$ comes from requiring that accretion is not
centrifugally inhibited at the minimum $\dot M$ at which coherent
pulsations were detected.  The lower limit on $\mu$ comes from requiring
that the accretion disk is terminated above the stellar surface at the
maximum $\dot M$ at which coherent pulsations were detected.  Note that
$\gamma_{\rm B}$ depends on $\dot M$, and so must have different values at
the minimum and maximum $\dot M$ considered here.  Conservatively,
$0.01\lesssim \gamma_{\rm B} < 1$.}
 \vspace*{0.7cm} 
 \end{minipage}
 \end{figure}

Finally, we can estimate the mass accretion rate $\dot{M}_{\rm eq}$ for
which the observed spin frequency of the neutron star is the equilibrium
frequency (at which the accretion torque is zero). The result is
 \begin{equation}
 \dot{M}_{\rm eq}=2\times 10^{-11}\omega_{\rm c}^{-7/3}
 \left(\frac{\gamma_{\rm B}}{0.1}\right)
 \left(\frac{\mu}{10^{26}~\mbox{G\,cm}^{3}}\right)^2
 \left(\frac{M}{1.4 M_\odot}\right)^{-5/3}
 \left(\frac{\nu_{\rm s}}{401~\mbox{Hz}}\right)^{7/3}
 M_\odot\;{\rm yr}^{-1}\;,
 \end{equation}
 where $\omega_{\rm c}$ is the critical fastness parameter (Ghosh \& Lamb
1979b). The value of $\dot{M}_{\rm eq}$ depends sensitively on the
(unknown) magnetic field strength and on $\gamma_{\rm B}(\dot M)$.
However, for values of these parameters consistent with the constraints
imposed above, $\dot{M}_{\rm eq}$ agrees with the long-term average
accretion rate of \sax\ inferred by Chakrabarty \& Morgan (1998) from the
fluence of the X-ray outburst and the outburst recurrence time.  Moreover,
in order for an LMXB with a 2~hr binary period to be below the separatrix
between transient and persistent systems in the diagram of donor mass
versus orbital period, it must have $\dot M\lesssim 3\times
10^{-11}~M_\odot$~yr$ ^{-1}$ (van Paradijs\markcite{P96} 1996; King, Kolb,
\& Szuszkewicz\markcite{KKS97} 1997), also consistent with the value
estimated above (Chakrabarty \& Morgan 1998). 

\section{DISCUSSION}

In this paper we have studied the disk-magnetosphere interaction in the
first known accretion-powered, millisecond pulsar, \sax.  We have
demonstrated that various simple models of inner accretion disks are not
consistent with observations.  We have also argued that the magnetic field
strength of \sax\ is a few times $10^8$~G at the stellar pole, using very
general constraints on the properties of the neutron star and on the
physics of the inner accretion disk flow. In this section, we discuss the
implications of our results for models of the disk-magnetosphere
interaction around weakly magnetic neutron stars.  We also compare \sax\
to the recycled millisecond radio pulsars and to the non-pulsing
neutron-star LMXBs. 

\subsection{The Disk-Magnetosphere Interaction in Weakly Magnetic
Neutron Stars} 

BATSE observations of strongly magnetic ($B\sim 10^{12}$~G)
accretion-powered pulsars in Be/X-ray transients have shown that the
scaling of the radius $r_0$ of interaction between the stellar magnetic
field and a gas-pressure dominated disk flow can account for the
observations (Ghosh 1996; Finger, Wilson, \& Harmon\markcite{FWH96} 1996;
Bildsten et al.\ 1997).  However, in \S2 we showed that this same scaling
is inconsistent with the detection of coherent pulsations throughout the
1998 April/May outburst of \sax, for {\em most\/} equations of state of
neutron star matter.  This is not surprising, given the very different
physical conditions in the interaction regions around weakly- and
strongly-magnetic neutron stars. 

In writing the stress balance equation~(\ref{bal}) and the
scaling~(\ref{gen}), we made a number of implicit assumptions
regarding the properties of the neutron star and the inner accretion
disk flow: first, that the stellar magnetic field is
dipolar; second, that the pulsar magnetic moment is parallel to its
spin axis; third, that the dominant mechanism for removing angular
momentum from the accreting gas in the disk is magnetic
stress; and finally, that the gravitational field is
Newtonian everywhere and hence that stable, circular Keplerian orbits
exist at all radii.  Although these assumptions are valid (or at least
are reasonable approximations) in the case of a strongly magnetic
neutron star, they break down when $r_0$ is comparable to the stellar
radius (see Lai\markcite{L98} 1998 for solutions of the structure of
the inner accretion disk where some of these effects have been taken
into account in a simple way).

Let us examine the validity of each of these assumptions.  First,
magnetic stresses would remove angular momentum faster than predicted
by equation~(\ref{gen}) if higher-order multipoles were present in the
pulsar magnetic field.  For example, if the stellar magnetic field
could be described entirely by a multipole of order $l$ (i.e, the
magnetic field strength in the equator was $B=\mu_l/r^{l+1}$), then
the Keplerian frequency at $r_0$ would be 
 \begin{equation}
 \nu_0=(2\pi)^{-1}\gamma_{\rm B}^{-3/(4l-1)}
 (GM)^{(2l+1)/(4l-1)}
 \mu^{-6/(4l-1)}
 \dot{M}^{3/(4l-1)}\;.
 \label{mult}
 \end{equation}
 Equation~(\ref{mult}) shows that, if $\gamma_{\rm B}$ depends only
weakly on $\dot{M}$, the dependence of $\nu_0$ on $\dot M$ weakens as
the order $l$ of the multipole increases: for a dipolar ($l=2$) field
$\nu_0\propto \dot{M}^{0.43}$, for a quadrupolar ($l=4$) field $\nu_0
\propto \dot{M}^{0.2}$, etc. As a result, the existence of intrinsic
or induced multipole moments in the magnetosphere can significantly
weaken the $\dot M$-dependence of $\nu_0$, hence making the
scaling~(\ref{mult}) consistent with observations of \sax.
As an illustration of this, Figure~3 shows the constraints imposed by
the 1G model on the mass and radius of the neutron star in \sax\ for
various contributions of an aligned quadrupole moment to the pulsar
magnetic field.  Clearly, even a modest quadrupole moment (which would
be completely negligible at distances larger than a few stellar
radii from the surface) is enough to make the 1G model consistent 
with observations.

\begin{figure}[t]
\begin{minipage}[b]{10.0cm}
\centerline{
\psfig{file=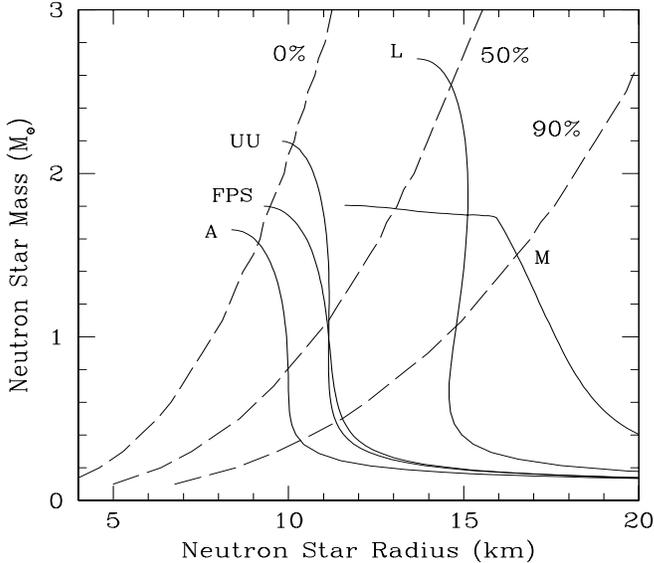,angle=0,height=7.8truecm,width=9truecm}}
\end{minipage}
\begin{minipage}[b]{6.0cm}
\figcaption[]
 {\footnotesize The effect of the presence of a quadrupole magnetic
moments on the constraints on the neutron star mass and radius in \sax.
The solid lines represent mass-radius relations of stable neutron stars
for various proposed equations of state (Cook et al. 1994).  The dashed
lines represent the limits imposed by the 1G model for the inner accretion
disk and for different fractional contributions of an aligned quadrupole
moment to the magnetic field of the neutron star. The allowed region for
each model is to the left of the corresponding dashed curve.}
 \vspace{1.4cm}
 \end{minipage}
\end{figure}

The intrinsic multipole moments in neutron star LMXBs are thought to be
weaker than their dipole moments, based on the current spin frequencies
and spin frequency derivatives of the millisecond radio pulsars thought to
be their descendants (Arons 1993).  However, electrical currents in the
accretion disk and the magnetosphere can significantly enhance the
strength of multipole moments by altering the magnetic field topology in
the interaction region (see, e.g., Ghosh \& Lamb 1979a).  Moreover,
electrical currents on the neutron star induced by the presence of an
accretion disk can produce multipoles as strong as the intrinsic dipole,
provided that the disk extends very close to the stellar surface (Psaltis,
Lamb, \& Zylstra\markcite{PLZ96} 1996).  The corotation radius $r_{\rm
co}=(GM/4\pi^2\nu_{\rm s}^2)^{1/3}$ (which is an upper bound on $r_0$ when
coherent pulsations are detected; see, e.g., Ghosh \& Lamb 1979a) is only
$\simeq 2.8$ times the neutron star radius in \sax, and therefore none of
the above effects are negligible.

Second, if any of the magnetic moments of \sax\ is 
not aligned to the stellar rotation axis (which is actually a necessary 
condition for the system to appear as a pulsar) then the scaling of the 
inner Keplerian disk radius with accretion rate may be different 
than what is predicted by the models considered above.  This case 
has never been treated in the literature because it corresponds to a 
time-dependent non-steady-state problem.  It is beyond the scope 
of the current paper to generalize models of the interaction between the 
pulsar magnetic field and the accretion disk to time-dependent 
situations. We will therefore not consider this case any further.

Third, when the accretion disk penetrates very close to the neutron star,
the effect of magnetic stresses is amplified by radiation drag forces that
can efficiently remove angular momentum from the accreting gas (Miller \&
Lamb 1996; Miller et al.\ 1998).  Finally, the specific angular momentum
of gas at radii comparable to the innermost stable orbit around a compact
object has a weaker dependence on radius than estimated in a Newtonian
gravitational field.  

All the effects discussed above result in a weaker $\dot M$-dependence
of $\nu_0$, thus making scaling~(\ref{gen}) consistent with
observations of \sax. 
 
\begin{figure}[t]
\begin{minipage}[b]{10.0cm}
\centerline{
\psfig{file=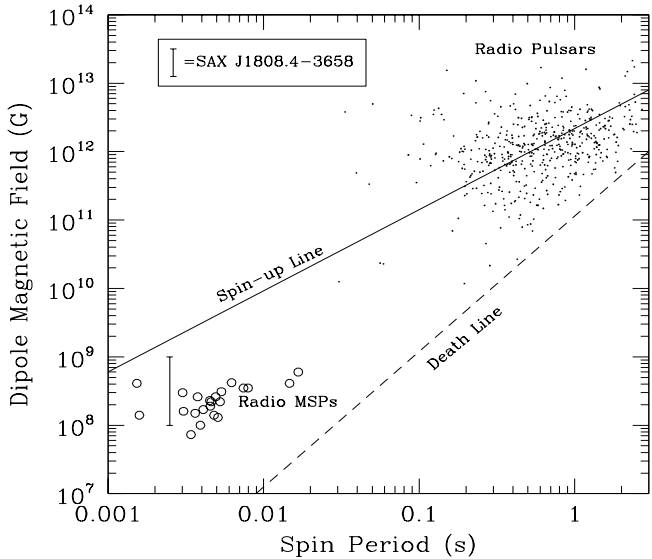,angle=0,height=7.8truecm,width=9truecm}}
\vspace{0.5cm}
\end{minipage}
\begin{minipage}[b]{6.0cm}
\figcaption[]
 {\footnotesize Inferred pulsar dipole magnetic field strength at the
stellar equator, as a function of spin period.  The small dots are normal
radio pulsars (Taylor et al.\ 1993).  The open circles are millisecond
radio pulsars which are not members of globular clusters, and for which
the inferred field strengths have been corrected for the apparent period
derivative caused by their transverse velocities (Camilo et al.\ 1994).  
The error bar shows the allowed range of field strengths for \sax.  The
spin-up line ({\em solid\/}) is shown here for illustrative reasons only
and represents the minimum period of recycled pulsars for a given field
strength, assuming a specific model for the inner accretion disk and the
disk-magnetosphere interaction.  The death line ({\em dashed\/}) is an
empirical estimate of the maximum period at which pulsars have detectable
radio emission for a given field strength.}
 \end{minipage}
\end{figure}

\subsection{Comparison with Millisecond Radio Pulsars}

When mass transfer onto the neutron star ends, \sax\ may appear as a
millisecond radio pulsar (Wijnands \& van der Klis 1998a; see also
Bhattacharya \& van den Heuvel\markcite{BH91} 1991). Figure~3 compares the
current spin frequency and inferred magnetic field strength of \sax\ at
the stellar equator with those of the known normal and millisecond radio
pulsars (Taylor, Manchester, \& Lyne\markcite{TML93} 1993).  We only
include millisecond radio pulsars in the Galactic disk, since the inferred
spin frequency derivatives (and hence the magnetic field strengths) of
pulsars in globular clusters are probably contaminated by gravitational
acceleration in the potential of the cluster.  Moreover, we use magnetic
field strengths that have been corrected for the effect of the apparent
spin frequency derivative caused by the non-negligible transverse
velocities of the pulsars (Camilo, Thorsett, \& Kulkarni\markcite{CTK94}
1994; Toscano et al.\markcite{Tosca99} 1999).  Note, however, that the
inferred magnetic dipole moments assume orthogonal dipole rotators,
and model-dependent systematic effects have not been taken into
account.  

The mass of the companion of \sax\ is $\lesssim 0.3 M_\odot$, and hence
the binary must be close to the termination of its X-ray phase
(Chakrabarty \& Morgan 1998). The current magnetic field strength and spin
frequency of the neutron star should thus be very similar to that of the
descendant millisecond radio pulsar.  Figure~3 shows that the descendant
of \sax\ should appear as a normal millisecond radio pulsar in the
Galactic disk. Given the small number statistics and the uncertainty in
our estimate of the magnetic field strength in \sax\ and in the millisecond
radio pulsars, it is not possible to determine whether the field strength
of \sax\ is lower or higher than those in the known millisecond radio
pulsars. Such a determination would be of interest in comparing \sax\ to
the other non-pulsing LMXBs. 

\subsection{Comparison with Non-Pulsing LMXBs}

\sax\ is the only known weakly-magnetic accreting neutron star with
persistent coherent pulsations at the neutron
star spin frequency. Upper limits on the amplitudes of periodic
oscillations in other neutron-star LMXBs range from $\lesssim 1$\% for the
most luminous Z sources to a few percent for the less luminous atoll
sources (see Vaughan et al.\markcite{Vetal} 1994 and references therein). 
Coherent millisecond oscillations, however, possibly at the spin
frequencies of the neutron stars, have been observed in several
neutron-star LMXBs during Type~I X-ray bursts (Strohmayer, Swank, \&
Zhang\markcite{Setal98} 1998).  

The absence of persistent coherent pulsations in most LMXBs might be
due to neutron star magnetic field strengths that are significantly
different than that in \sax.  Specifically, the field strengths in the
non-pulsing LMXBs may be so strong that accretion is centrifugally
inhibited, or so weak that the accretion disk extends all the way to
the neutron star surface.  Alternatively, the neutron stars in the
non-pulsing LMXBs may be surrounded by optically thick scattering
media that spread the pulsar beams and attenuate the oscillations
below present detection thresholds (Brainerd \& Lamb 1987; Kylafis \&
Phinney 1989). We consider here each possibility separately.

It is rather unlikely that the field strengths of all non-pulsing
LMXBs are so strong that accretion is centrifugally inhibited, since
at least some are accreting at near-Eddington rates.  Moreover, if the field
strengths were all so strong, then the observational fact that all
known millisecond radio pulsars lie below the so-called spin-up line
in the $B$-$P$ diagram (Fig.~3) would have to be a selection effect or
a coincidence.  This is because the spin-up line defines the minimum
spin period of a recycled pulsar (which is the equilibrium period at
the Eddington accretion rate) for a given field strength, provided
that accretion is not centrifugally inhibited.

On the other hand, if the field strengths in the non-pulsing LMXBs are
significantly weaker than that of \sax, then a significant fraction of
the gas in the accretion disk may be reaching the stellar surface in
the disk plane, producing undetectable periodic oscillations at the
neutron star spin frequency. This is apparently consistent with
identifying the frequencies of the observed kHz QPOs in the
non-pulsing LMXBs with Keplerian orbital frequencies in the inner
accretion disk (van der Klis et al.\ 1996; Strohmayer et al.\ 1996;
Miller et al.\ 1998). 

However, there are at least two other transient LMXBs
(Aql~X-1 and 4U~1608$-$52) whose outburst histories, X-ray spectral
properties, and broad-band noise features are similar to those of \sax\
(see, e.g., Heindl \& Smith 1998; Wijnands \& van der Klis 1998b), and
yet do not show periodic pulsations in their persistent emission.
At the same time, these systems do show kHz QPOs and have other timing
and spectral properties characteristic of the non-pulsing LMXBs
(Zhang, Yu, \& Zhang\markcite{ZYZ98b} 1998b; Yu et al.\markcite{Yetal97}
1997).  In particular, the X-ray spectrum of 4U~1608$-$52, which is a direct
probe of the structure and dynamics of the inner accretion disk flow,
is nearly identical to that of \sax\ (Heindl \& Smith 1998), with the
same power-law index and high-energy cut-off.  It seems unlikely that
the X-ray spectra of these two sources could be so similar if their
field strengths (and thus their inner disk radii) differ
significantly.  This is particularly true of the spectra at high
energies (20--100~keV), which must be formed close to the neutron star
surface and are thus highly sensitive to the inner disk flow geometry.
In fact, if the high-energy spectra are produced by Comptonization of
high-harmonic self-absorbed cyclotron photons (Psaltis et al.\ 1995),
then the stellar field strengths can be inferred from the hardness of
the $2-20$~keV X-ray spectra (Psaltis\markcite{P98} 1998; Psaltis \&
Lamb 1998).  Indeed, the magnetic field strengths of the
low-luminosity atoll sources, such as Aql~X-1 and 4U~1608$-$52, have
been inferred to be $\sim 5\times 10^8$~G (see Psaltis \& Lamb 1998),
i.e., very similar to the magnetic field strength of SAX~J1808$-$36 we
estimated in \S3. 

If the field strength in \sax\ is comparable to that of some
non-pulsing LMXBs, then the absence of periodic oscillations in the
persistent emission from most LMXBs must be attributed to another,
external effect.  The presence of a scattering medium around the
neutron star and its magnetosphere has been invoked previously as an
explanation of the non-detection of 
periodic oscillations in neutron-star LMXBs (Brainerd \& Lamb 1987;
Kylafis \& Phinney 1989). The high ($\tau\sim 4-5$) optical depths
inferred from X-ray spectra of the Z and atoll sources are consistent
with this hypothesis (see Psaltis et al.\ 1995 and references
therein). However, {\em RXTE}/HEXTE observations of \sax\ indicate 
an optical depth of $\tau=4.0\pm 0.2$, based on the slope and
high-energy cutoff of the X-ray spectrum, at the same time that
pulsations were detected (Heindl \& Smith 1998; Heindl 1998,
private communication; note, however, that Gilfanov et al.\ [1998]
analyzed the same data and reached a different conclusion.).  This is
very similar to the optical depth inferred for the non-pulsing source
4U~1608$-$52 (Heindl \& Smith 1998), which we argued above has a
comparable magnetic field strength.  Thus, if the optical depth of
\sax\ is indeed $\gtrsim 3$, then the reason coherent pulsations are
detected only from this LMXB cannot simply be weaker attenuation of
the beaming oscillation in the scattering medium.

A number of additional effects may be responsible for the detectability of
coherent pulsations in \sax.  For example, its very small measured mass
function is suggestive of a nearly face-on viewing angle of the binary
(Chakrabarty \& Morgan 1998).  In this case, the pulsed emission may
propagate through a region of lower optical depth close to the pulsar
rotation poles, whereas the majority of the (non-pulsed) X-ray photons
may travel through optically thick regions nearby.  However, it is important
to note that the presence of a shallow partial eclipse in \sax\ has
been suggested (Chakrabarty \& Morgan 1998; Heindl \& Smith 1998).
Confirmation of this feature would rule out a face-on orientation of
the binary, thus challenging the above explanation.  Alternatively,
different stellar field topologies in \sax\ and the non-pulsing
neutron star LMXBs, possibly related to different prior evolution or
even to different magnetic inclinations, may be responsible. 

A number of other LMXBs share many characteristics of \sax\ and are
thus good candidates for detecting periodic pulsations in their
persistent emission. For example, as we discussed above, Aql~X-1 and
4U~1608$-$52 are transients with X-ray spectra very similar to \sax,
suggesting similar field strengths and inner disk flows. The nearly
sinusoidal optical orbital-phase lightcurves of three other LMXBs
(4U~1636$-$53, GX~9$+$9, and 4U~1735$-$44) suggest that these are also
viewed nearly face-on (see, van Paradijs \& McClintock\markcite{PM}
1995).  Finally, GS~1826$-$26 (Homer, Charles, \&
O'Donoghue\markcite{HCO98} 1998) and MS~1603$+$2600 (Hakala et
al.\markcite{Hetal98} 1998) are other LMXBs with low-mass companions
in $\sim 2$-hour orbits, in which mass transfer is probably driven by
angular momentum losses due to gravitational radiation.  These
binaries are thus in an evolutionary stage very similar to that of
\sax.  In particular, GS~1826$-$26 is a transient X-ray burster with a
low-amplitude modulation of its optical brightness, again possibly
suggesting a face-on orientation (Homer et al.\ 1998).  Detecting or
imposing stringent upper limits on coherent pulsations in the
persistent emission of any of these sources will be crucial in
understanding the characteristics of \sax\ and the relationship
between LMXBs and millisecond radio pulsars.
 
\acknowledgments We are grateful to W.\ Heindl and Wei Cui for sharing
their results prior to publication. We also thank Steve Thorsett for
pointing out some recent work on the magnetic fields of recycled pulsars,
Uli Kolb for discussions on the disk-instability model of X-ray transients
and on the binary properties of \sax, as well as Lars Bildsten, Vicky
Kalogera, Michiel van der Klis, and Rudy Wijnands for many useful
discussions and for carefully reading the manuscript. This work was
supported in part by a postdoctoral fellowship of the Smithsonian
Institution (D.\,P.), by a NASA Compton {\em GRO\/} Postdoctoral
Fellowship at MIT under grant NAG~5-3109 (D.\,C.), and by several NASA
grants under the {\em RXTE\,} Guest Observer Program.  We are also
grateful for the hospitality of the Astronomy Group at the University of
Leicester and the Astronomical Institute at the University of Amsterdam,
where parts of this work were completed.

\clearpage

{\small
\begin{deluxetable}{ccccc}
\tablewidth{350pt}
\tablecaption{Scaling of the Keplerian Frequency at
the Coupling Radius\tablenotemark{a}}
\tablehead{Disk Model\tablenotemark{b} & $\nu_{\rm 
K,0}$~(Hz) & $\alpha$ & $\beta$ & $\gamma$}
\startdata
1G & 430 & 0.38 & $-$0.87 & 0.82\\
1R & 210 & 0.23 & $-$0.77 & 0.70\\
2B & 80 & 0.72 & $-$0.86 & 0.43\\
2S & 50 & 2.55 & $-$1.20 & $-$0.60\\
\enddata
\tablenotetext{a}{After Ghosh \& Lamb 1992.}
\tablenotetext{b}{1G: Optically thick, gas-pressure
dominated (GPD) disk; 1R: Optically thick,
radiation-pressure dominated (RPD) disk; 2B: 
Two-temperature, optically thin GPD disk with
Comptonized bremsstrahlung; 2S: Two-temperature,
optically thin GPD disk with Comptonized soft photons
(for references see Ghosh \& Lamb 1992).}
\end{deluxetable}
}

\end{document}